\documentclass[12pt]{iopart}

\def\be{\begin{equation}}
\def\te{\end{equation}}
\def\ee{\end{equation}}
\def\ba{\begin{eqnarray}}
\def\bea{\begin{eqnarray}}
\def\nn{\nonumber\\}
\def\tea{\end{eqnarray}}
\def\ea{\end{eqnarray}}
\def\eea{\end{eqnarray}}

\begin{document}


%
%

\title{Decoherence in the cosmic background radiation}

\author{Mariano Franco$^1$ and Esteban Calzetta$^{2}$}
\address{$^1$ Departamento de F\'isica, FCEyN-UBA, Ciudad Universitaria, Pabell\'on I, Buenos Aires, 1428, Argentina}
\address{$^2$ Departamento de F\'isica, FCEyN-UBA and IFIBA-CONICET, Ciudad Universitaria, Pabell\'on I, Buenos Aires, 1428, Argentina}
\eads{\mailto{mfranco@df.uba.ar}, \mailto{calzetta@df.uba.ar}}

\begin{abstract}
In this paper we analyze the possibility of detecting nontrivial quantum phenomena in observations of the temperature anisotropy of the cosmic background radiation (CBR), 
for example, if the Universe could be found in a coherent superposition of two states corresponding to different CBR temperatures. 
Such observations are sensitive to scalar primordial fluctuations but insensitive to tensor fluctuations, which are therefore converted into an environment 
for the former. Even for a free inflaton field minimally coupled to gravity, scalar-tensor interactions induce enough decoherence among histories of the scalar fluctuations 
as to render them classical under any realistic probe of their amplitudes.
\end{abstract}


\section{Introduction}
According to the inflationary paradigm \cite{Wei,Bass}, not only primordial cosmological fluctuations are quantum in origin, but they are also created in a very non-classical state \cite{CamPar06,CamPar08a,CamPar08b,MazHey09}. This raises the tantalizing possibility of uncovering nontrivial quantum behavior through cosmological observations. However, no known cosmological probe would reveal the actual quantum state of primordial fluctuations since all known methods of observation focus on a restricted set of properties of those fluctuations, thus leaving a remainder which must be traced over. Therefore, to discuss nontrivial quantum behavior we have to consider not only the quantum features of the cosmological fluctuations, but also the loss of quantum coherence induced by the partial description appropriate to the observation in question.

In this paper we take as example the observations of the temperature anisotropy amplitudes of the cosmic background radiation (CBR). The temperature fluctuations are determined by the scalar cosmological fluctuations. Unlike the case when CBR polarization is being observed,   tensor perturbations affect the result only through their action on the scalar ones. Therefore in the observation of CBR temperature fluctuation amplitudes, we must regard tensor fluctuations as an environment coupled to the system of interest, namely the scalar fluctuations.

The coupling between the system and its environment induces decoherence in the former \cite{JZKG,Schl}. 
Adopting the Hartle- Gell-Mann consistent histories approach to quantum mechanics \cite{Gell-Har,Har}, we ask whether it is possible to observe the coherence between different histories of the scalar fluctuations, after tracing over the tensor fluctuations. We shall only consider the coupling between these fluctuations demanded by general relativity. We represent all matter fields by a single free scalar inflaton field, minimally coupled to gravity. After identifying the relevant gauge invariant variables and imposing the Newtonian gauge conditions (see below), the momentum constraints of general relativity relate the inflaton field to the single scalar degree of freedom in the metric. Thus, there is only one gauge invariant scalar degree of freedom in the theory. This scalar field is coupled to the graviton field, which after making the graviton polarization explicit may also be described by two scalar fields. We disregard vector perturbations. 

Our conclusion is that the decoherence induced by tensor perturbations is strong enough to erase any traces of quantum behavior in the scalar fluctuations, given any realistic observational scenario by today's standards. To this extent, our findings are consistent with other treatments of the issue in the literature. These other approaches are based either on different system-environment splits or on averaging over the decaying mode of the cosmological fluctuations \cite{GP,CalHu95,CalGon97,KPS,KLPS,KP,Lom-Lop,ProRig07,Bur,Mar}.

Within the Gell-Mann and Hartle formalism one has the freedom to take any pair of histories to compute the decoherence functional. We choose these histories to see whether quantum effects in the CBR spectrum can be detected. According to the present paradigm, the amplitudes of the temperature fluctuations in the different modes in which the CBR may be decomposed are the result of a stochastic process. The amplitudes themselves are independent very nearly gaussian random variables. We regard each realization of this process as a ``history'' and ask whether decoherence between different, independent typical histories may be observed. Since the histories themselves are random, we shall compute the expectation value of the influence functional between two independent histories. We will also show that the mean quadratic deviation of the influence functional from its expectation value is negligible.

To traslate the instantaneous picture of the CBR temperature fluctuations at the time of last scattering into a history of scalar fluctuations evolving in space-time, we  use the Sachs-Wolfe effect \cite{Wei,Sachs,Dod,Lin}. This allows us to find the amplitudes of the growing modes in the scalar fluctuations corresponding to given temperature fluctuations. To link the amplitude of scalar perturbations in the recombination era with the inflationary period we use Bardeen's conservation law \cite{Bard,MFB}. Once we have associated a history of the scalar fluctuations to the given temperature fluctuations, we compute the expectation value and the standard deviation of the decoherence induced by the gravitons on two independent histories chosen at random. 

This paper is organized as follows. In Section \ref{cosmpert} we give a brief summary of inflation, gauge invariant cosmological perturbations, their link to CBR temperature and we compute the interaction between the scalar and tensor modes which is necessary to calculate the decoherence functional. Section \ref{dec} is devoted to decoherence: we first give a brief summary of the Hartle- Gell-Mann approach (mainly to fix our notation) and and then we compute the decoherence functional and its standard deviation. Finally, Section \ref{conc} contains our conclusions.

\section{Inflation and cosmological perturbations} \label{cosmpert}

The aim of this work is to compute the decoherence suffered by the scalar perturbations due to its interaction with the tensor perturbations in the inflationary stage of the Universe. For such calculation it is necessary to find the interaction between the perturbations. In this section we calculate the interaction between scalar and tensor modes using the ADM formulation of General Relativity \cite{MTW}. Then, we compute the free action of the tensor perturbations and the Hadamard propagator \cite{CalHu} associated to them. We will use the \textit{Newtonian gauge} to find the invariants cosmological perturbations \cite{MFB}. 

Let us begin by describing the cosmological model we have in mind. We shall adopt natural units in which $c=\hbar=k_B=1$ and therefore the Plank mass $m_{pl}=10^{19}\:\mathrm{GeV}$. The present temperature of the Universe is $T_0=10^{-13}\:\mathrm{GeV}$ and the present age of the Universe (which is also essentially the size of the present day cosmological horizon) is $L_0=10^{42}\:\mathrm{GeV}^{-1}$. Up to that distance the Universe is well described by a spatially flat Friedmann - Robertson - Walker (FRW) model with a scale factor $a\left(t\right)$; we assume $a=1$ at the present time. We also assume that the Universe underwent a stage of inflationary expansion which ended at the time of reheating $t_r$. For concreteness we assume a reheating temperature of $T_r=10^{16}\:\mathrm{GeV}$. This means that at the time of reheating, and therefore also during the inflationary era, the Hubble parameter was $H=T_r^2/m_p=10^{13}\:\mathrm{GeV}$. The scale factor at reheating was $a_r=T_0/T_r=10^{-29}$. In terms of conformal time $\eta=-1/aH$ this means inflation ends at a time $\left|\eta_r\right|=10^{16}\:\mathrm{GeV}^{-1}$. This represents that our present horizon crossed the horizon during inflation at the time when the conformal factor was $a_i=1/HL_0=10^{-55}$. Observe that as expected $a_r/a_i=10^{26}=e^{60}$. At that moment the conformal time was $\left|\eta_i\right|=10^{42}\:\mathrm{GeV}^{-1}$. For all practical purposes, we shall take this event as the beginning of inflation. We shall be concerned with cosmological modes which crossed the horizon during inflation sometime between $\left|\eta_i\right|$ and  $\left|\eta_r\right|$. This means their conformal wave numbers are in the range $10^{-42}\:\mathrm{GeV}<q<10^{-16}\:\mathrm{GeV}$. Concretely, the mode $q$ crosses the horizon at a conformal time $\left|\eta_e\right|=1/q$, when $1/a\left(\eta_e\right)H=1/q$.

\subsection{Quick review of Inflation}  

The necessary condition to achieve an accelerated expansion is $p=-\rho$. This condition yield the De-Sitter stage when the scale factor grows exponentially, $a\sim e^{Ht}$ \cite{Wei,Bass,Guth}. This stage of evolution is dominated by a homogeneous scalar field called inflaton ($\varphi_0$). Its energy density and pressure are given by
\numparts
\begin{eqnarray}
\rho&=\frac{1}{2}\dot{\varphi}^2_0+V(\varphi_0)\\
p&=\frac{1}{2}\dot{\varphi}^2_0-V(\varphi_0)
\end{eqnarray}
\endnumparts 
where $V(\varphi_0)$ is the potential energy of the inflaton.

In an expanding, homogeneous and isotropic space-time described for the plane FRW metric - $ds^2=-dt^2+a^2(t)dx^2$ - the inflaton follows the field equations,
\numparts
\begin{eqnarray}
H^{2}&= \frac{8 \pi}{3m^2_{pl}}\big[\frac{1}{2}\dot{\varphi}^{2}_0+V(\varphi_0)\big]\label{10}\\
0&=\ddot{\varphi}_0+3H\dot{\varphi}_0+\frac{\partial V}{\partial \varphi_0} \label{11}
\end{eqnarray}
\endnumparts 
where $H=\dot{a}/a$ is the Hubble factor (approximately constant during inflation) and $m_{pl}$ is the Planck mass. 

The inflationary condition requires a sufficiently flat potential so that the potential energy dominates over the kinetic energy, $\dot{\varphi}^2_0<V(\varphi_0)$. This condition, known as slow-roll, is satisfied if
\numparts
\begin{eqnarray}
 \epsilon=\frac{m^2_{pl}}{16 \pi}\left( \frac{V_{,\varphi}}{V}\right)^2 <<1 \label{slowroll}\\
\zeta=\frac{m^2_{pl}}{8 \pi}\frac{V_{,\varphi \varphi}}{V}<<1
\end{eqnarray}
\endnumparts 
where $\epsilon$ and $\zeta$ are the so-called slow-roll parameters. 

Using equation (\ref{slowroll}) to rewrite $V_{,\varphi}$ in terms of $\epsilon$ and neglecting the $\ddot{\varphi}_0$ term in (\ref{10}), the first slow roll parameter can be written in terms of the kinetic and potential energies as

\begin{equation}
\epsilon \approx \frac{\dot{\varphi}^2_0}{V} \label{epsilon}
\end{equation} 
Now, using that $V=m^2_{\varphi_0}\varphi^2$ in (\ref{slowroll}) the inflaton field results

\begin{equation}
 \varphi_0=\frac{m_{pl}}{\sqrt{\epsilon}} \label{inflaton}
\end{equation}
and the Hubble factor is

\begin{equation}
 H\sim \frac{m_{\varphi}}{\sqrt{\epsilon}} \label{Hminflaton}
\end{equation} 
It will be convenient to put the time derivative of the inflaton field ($\dot{\varphi}_0$) in terms of the conformal time $\eta$. A derivative with respect to $\eta$ is denoted by $f'$. We also define $\mathcal{H}=a'/a=aH$.

Thus, using the conformal time and the slow-roll parameters, equation (\ref{11}) becomes

\begin{equation}
 \varphi'_0\approx\sqrt{\epsilon}\frac{m_{pl}}{\eta} \label{inflatonderivative}
\end{equation}  
We will use those equations in the next subsections and in Section \ref{dec} in order to compute the decoherence functional.

\subsection{Invariant cosmological perturbations}

Perfectly homogeneous and isotropic space-time is only an idealization. This description cannot explain the large structures observed in the Universe. One way to achieve a satisfactory explanation for the structure distribution is to include small perturbations in the FRW metric.

We will consider only linear perturbations about the fields,

\begin{equation}
 \zeta=\zeta_0(t)+\delta\zeta(t,x)
\end{equation} 
The linear part of the perturbed FRW metric is \cite{MFB},

\begin{eqnarray}
ds^{2}&=
a^{2}(\eta)\left\lbrace
(1+2\phi)d\eta^{2}-2(S_i+B_{;i})dx^{i}d\eta 
\right.\nonumber\\
&\left.
-\left[(1-2\psi)\gamma_{ij}
+F_{i;j}+F_{j;i}+2E_{;ij}+h_{ij}\right]dx^{i}dx^{j}\right\rbrace \label{pertur}
\end{eqnarray} 
where the $;$ sub index is the covariant derivative respect to the background space-time $\gamma_{ij}$. In the flat FRW space-time $\gamma_{ij}=\delta_{ij}$ and therefore the covariant derivative is the usual one.

The perturbations can be split into scalar, vector and tensor components according to their transformation properties  in the spatial hyper surfaces. The scalar perturbations are $\phi$, $B$, $\psi$ and $E$. 

The vector component is given by $S$ and $F$ which satisfy $S^{;i}_i=F^{;i}_i=0$. The symmetric tensor $h_{ij}$ gives tensor perturbations with the constraints $h^i_i=0$ and $h^{;j}_{ij}=0$.

All those variables are gauge dependent. To describe the inhomogeneities of the universe through linear perturbations, we must first distinguish which of the quantities have a well defined physical interpretation and is not related to a change of coordinates or a change in the system of reference. There is an infinite number of invariant quantities, but two commonly used are \cite{MFB}
\numparts
\begin{eqnarray}
 \Phi&=\phi+\frac{1}{a}[(B-E')a]'\\
 \Psi&=\psi-\frac{a'}{a}(B-E')
\end{eqnarray}
\endnumparts 

The reason for choosing these quantities is that in the \textit{Newtonian
gauge}, where $B=E=0$, the two gauge invariant quantities coincide with the scalar perturbations in the metric, $\Phi=\phi$ and $\Psi=\psi$. Moreover, when the spatial part in the perturbation of the energy moment tensor is diagonal, the scalar perturbations $\phi$ and $\psi$ are equal and only one scalar degree of freedom in the metric remains. Furthermore,  a scalar quantity that is not included in the metric is already gauge invariant.

Regarding the tensor perturbations, they are gauge invariant by definition. Having zero trace and divergence, they do not have quantities that transform as scalars or vectors.

The ADM parameterization of the metric in terms of gauge invariant variables is as follows.
The shift function is

\begin{equation}
 N_i=a^2B_{,i}
\end{equation} 
the lapse function is
\begin{eqnarray}
N^2&-N_iN^i=a^2\left(1+2\phi\right)\nonumber\\
N^2&=a^2\left(1+2\phi+B_{,i}B_{,i}\right)\nonumber\\
N&\approx a\left(1+\phi-\frac{1}{2}\phi^2\right)
\end{eqnarray} 
and the extrinsic curvature tensor is

\begin{eqnarray}
\fl K_{ij}=& a(\eta)\left\lbrace (1-\phi)B_{ij}- ^{(3)}\Gamma^k_{ij}B_{,k}  \left[\phi'\left(1-2\phi\right)\mathcal{H}-\phi\phi'+\phi\left(1-2\phi\right)\mathcal{H}
\frac{3}{2}\phi^2\mathcal{H}\right]\delta_{ij} \right.\nonumber\\
\fl &\left.+\left(-1+\phi-\frac{3}{2}\phi^2\right)\mathcal{H}h_{ij}+\frac{1}{2}\left(-1+\phi-\frac{3}{4}\phi^2\right)h'_{ij}\right\rbrace 
\end{eqnarray}
where 

\begin{equation}
 \Gamma^k_{ij}=\frac{1}{2}g^{kl}\left( g_{il,j}+g_{jl,i}-g_{ij,l} \right)
\end{equation}
is the spatial part of Christoffel's coefficients with 

\begin{equation}
g_{ij}=-a^2(\eta)\left[(1-2\phi)\delta_{ij}+h_{ij}\right]
\end{equation}

being the spatial part of the plane perturbed metric without vector perturbations. 

So far, we have defined the scalar perturbation in the \textit{Newtonian gauge}, now we move on to analyse its dynamics and its link to CBR temperature.

\subsection{Free scalar perturbations and CBR temperature}\label{spCBR}

The evolution of $\phi$ is obtained from the perturbed Einstein's equations. Let us write  $u=(a/\varphi'_0)\phi$. Under the slow-roll approximation $u$ obeys the equation \cite{MFB}

\begin{equation}
 u''-\nabla^2 u - \frac 2{\eta^2}u=0
\end{equation} 
The equation for the modes $u_\mathbf{k}$ results in

\begin{equation}
 u''_\mathbf{k}(\eta)+(k^2- \frac 2{\eta^2})u_\mathbf{k}(\eta)=0
\label{pi0}
\end{equation} 
As $\left|\eta\right|\to 0$, we see there is a \emph{growing} mode and a \emph{decaying} mode. The latter becomes negligible against the former and is the sole contribution to CBR temperature fluctuations. We shall assume the $\phi$ field is a superposition of growing modes only, namely

\begin{equation}
\phi(\mathbf{x},\eta)=\int \frac{d^3k}{\left(2\pi\right)^3} e^{i\mathbf{kx}}\phi_{\mathbf{k}} F_k\left(\eta\right)
\label{pi1}
\end{equation}
where

\be
F_k\left(\eta\right)=\cos\left(k\eta\right)+k\eta\sin\left(k\eta\right)
\label{pi2}
\te
It is readily seen that $F_k/\eta$ is a solution to equation. (\ref{pi0})

Once the modes cross the horizon, $k\sim a_eH$, their amplitudes  are frozen at the value $\phi_{\mathbf{k}}$ until they re-enter into the recombination era. At this stage their amplitudes are amplified and can be related to the inflationary stage through equation \cite{Bard,MFB}

\be
\phi_{\mathbf{k}}\approx\frac{\dot{\varphi}^2_0}{V(\varphi_0)}\phi_{\mathbf{k}}(\eta_{k,rec})
\label{pi3}
\te
where $\eta_{k,rec}$ is the $k$ dependent time of final horizon crossing. 
Moreover, using the Sachs-Wolfe effect \cite{Wei,Sachs,Dod,Lin} we can relate the scalar perturbation with anisotropies in the Cosmic Background Radiation during the recombination period as follows

\begin{equation}
\frac{\delta T_\mathbf{k}}{T_0}=\frac{1}{3}\frac{V(\varphi_0)}{\dot{\varphi}^2_0}\phi_{\mathbf{k}}\label{CBR}
\end{equation}

With this last equation we can relate the scalar perturbation modes during inflation with the CBR anisotropies, which are an observable magnitude.

\subsection{Scalar-Tensor interaction}
The scalar perturbation $\phi$ (\textit{in Newtonian gauge}) and the perturbation $\delta\varphi$ to the inflaton field $\varphi_0$ are linked through the equation \cite{MFB}

\begin{equation}
 \phi'+\mathcal{H}\phi=4\pi m^{-2}_{pl} \varphi'_{0}\delta\varphi
\end{equation} 
Then, a single scalar degree of freedom remains in the Newtonian gauge. 

We consider now the derivation of the coupling current between the gauge invariant scalar mode $\phi$ and the gravitons. We start with the usual Einstein-Hilbert action written in ADM form \cite{MTW} plus the Klein-Gordon action for the inflaton field

\begin{eqnarray}
 S&=\frac{m^2_{pl}}{2}\int\;d^4x\left[
Ng^{1/2}(K^i_jK^j_i-K^2)+ \frac{1}{2}(g^{1/2}g^{ij}N)_{,i}(ln\;g)_{,j}N_{,i}(g^{1/2}g^{ij})_{,j}-\right.  \nonumber\\
 &-\left.\frac{1}{2}g^{1/2}N ^{(3)}\Gamma^k_{ij}g^{ij}_{,k}\right] +\int\;d^4x \sqrt{-g}\left[ \frac{1}{2}g_{\mu\nu} \varphi^{;\mu} \varphi^{;\nu} -V(\varphi) \right] 
\end{eqnarray} 
where $2K_{ij}= N^{-1}[N_{i;j}+N_{j;i}-g'_{ij}]$ is the extrinsic curvature tensor, $N$ the lapse function and $N_{,i}$ the shift function.

The extrinsic curvature tensor does not contribute to the scalar-tensor coupling; neither do terms containing the trace $h_{ii}=0$. Keeping terms containing two scalar and one graviton field we get

\numparts
\begin{eqnarray}
\frac{1}{2}m^2_{pl}N_{,i}(g^{1/2}g^{ij})_{,j}&\mapsto -\frac{1}{2}m^2_{pl}a^2\phi_{,i}\phi_{,j}h_{ij} \label{inter1}\\
-\frac{1}{2}m^2_{pl}\frac{1}{2}Ng^{1/2}\Gamma^k_{ij}g^{ij}_{,k}&\mapsto -2m^2_{pl}a^2\phi_{,i}\phi_{,j}h_{ij} \label{inter2}\\
\frac{1}{2}m^2_{pl}\frac{1}{2}(g^{1/2}g^{ij}N)_{,i}\frac{g_{,j}}{g}&\mapsto 3m^2_{pl}a^2\phi_{,i}\phi_{,j}h_{ij} \label{inter3}\\
\frac{1}{2}Ng^{1/2}g^{ij}\varphi_{,i}\varphi_{,j}&\mapsto \frac{1}{2}a^2\delta\varphi_{,i}\delta\varphi_{,j}h_{ij} \label{inter4}
\end{eqnarray} 
\endnumparts
We may summarize these equations writing

\begin{equation}
S_{int}=\int d^4x J_{ij} h_{ij}
\label{Sint}
\end{equation}
where

\bea
\fl J_{ij}&=\frac{m^2_{pl}}{2} a^2(\eta) \left[ \left(
1+16m^2_{pl}\mathcal{H}^2\varphi'^{-2}_0\right) \phi_{,i}\phi_{,j}+16m^2_{pl}\varphi'^{-2}_0\phi'_{,i}\phi'_{,j}+32
m^2_{pl}\mathcal{H}\varphi'^{-2}_0\phi_{,i}\phi'_{,j}\right]\nn
\fl &\approx\frac{m^2_{pl}}{2} a^2(\eta)\phi_{,i}\phi_{,j}  \label{cc}
\tea
This is the coupling current that is used to calculate the decoherence induced on the scalar tensor modes. To this aim it is also necessary to calculate the Hadamard propagator: equation. (\ref{fd}) below shows that the decoherence functional is written in terms of the Hadamard propagator and to compute it requires first to find the free action for the tensor perturbations.

\subsection{Free graviton Hadamard propagator}

To second order in $h_{ij}$, the free action of the gravitons is the usual Klein-Gordon action for tensors $h_{ij}$  

\begin{equation}
 S_{free}=\frac{m^2_{pl}}{4}\int d^4x a^2(\eta)\left[
h'_{ij}h'_{ij}-h_{ij,k}h_{ij,k}\right] 
\end{equation} 
The free dynamics of the gravitons is described in terms of their physical degrees of freedom \cite{CalHu95}

\begin{equation}
 h_{ij}(x)=\frac{1}{a(\eta)m_{pl}}\int d^3\mathbf{y}
[G^+_{ij}(\mathbf{x-y})h^+(\eta,\mathbf{y})+(+\leftrightarrow \times)]
\end{equation}  
where $+$ and $\times$ are the graviton polarizations and

\begin{equation}
 G^+_{ij}(\mathbf{x-y})=\int \frac{d^3k}{(2\pi)^3}e^{i\mathbf{k}(\mathbf{x}-\mathbf{y})}A^{+}_{\mathbf{k}ij} \label{tp}
\end{equation} 
The matrix $A_{ij}$ verifies

\begin{equation}
 A^{+}_{\mathbf{k}ii}=\mathbf{k}_iA^{+}_{\mathbf{k}ij}=A^{\times }_{\mathbf{k}ii}=\mathbf{k}_i A^{\times }_{\mathbf{k}ij}=0 \label{tepol}
\end{equation} 
and $h(\eta, y)$ obeys

\begin{equation}
 h''+2\frac{a''}{a}h-\nabla^2h=0
\end{equation} 
We assume the scalar field $h(y)$ is in the usual Bunch-Davis vacuum state.

The scalar Hadamard propagator is defined as

\begin{equation}
 G_1(y,y')= \left\langle h(y)h(y')+h(y')h(y) \right\rangle
\end{equation} 
It results

\be
G_1\left(y,y'\right)= \int\frac{d^3k}{(2\pi)^3} e^{i\mathbf{k}\left(\mathbf{y}-\mathbf{y}'\right)}\frac1{k}G_{1k}\left(\eta,\eta'\right)
\label{aaa}
\te
where

\be
G_{1k}\left(\eta,\eta'\right)= 
(1+\frac{1}{k^2\eta\eta'})\cos[k(\eta-\eta')]
+\frac{1}{k}(\frac{1}{\eta}-\frac{1}{\eta'})\sin[k(\eta-\eta')] \label{hp}
\te 
Therefore for the gravitons themselves we get

\bea
G_{1ijlm}\left(x,x'\right)&=&\left\langle \left\{h_{ij}\left(x\right),h_{lm}\left(x'\right)\right\}\right\rangle\nn
&=&\frac{1}{a(\eta)a(\eta')m_{pl}^2}\int\frac{d^3k}{(2\pi)^3} \Delta_{\mathbf{k}ijlm}{e^{i\mathbf{k}(\mathbf{x}-\mathbf{x}')}}G_{1k}\left(\eta,\eta'\right) \label{ghp}
\tea
where

\be
\Delta_{\mathbf{k}ijlm}=A^{+}_{\mathbf{k}ij}A^{+}_{\mathbf{k}lm}+A^{\times}_{\mathbf{k}ij}A^{\times}_{\mathbf{k}lm}
\label{ghpp}
\te

In the next Section we compute the decoherence functional using the coupling current given by (\ref{cc}) and the Hadamard propagator showed in (\ref{hp}). But since we have related the scalar cosmological perturbations to CBR temperature fluctuations in Section \ref{spCBR}, this allows to us to rewrite the decoherence functional in terms of this observable and can analyze if it is possible to detect nontrivial quantum effects on it.

\section{Decoherence Functional} \label{dec}

We use the Hartle Gell-Mann formalism to quantify the decoherence induced by the gravitons on the scalar perturbations. We first give a brief description of this formalism and then we compute the decoherence.

\subsection{Hartle Gell-Mann formalism}

In this section we give a quantitative discussion of decoherence. To calculate the loss of coherence induced on the scalar tensor modes (which are in the FRW metric) we use the decoherence functional developed by Gell-Mann and Hartle \cite{Gell-Har,Har}. We give a brief description of closed quantum systems including the decoherence term that is related to the classical sum rule of probabilities for different histories of a closed quantum system.

In the consistent histories description there is a subset of configuration space variables that are distinguished ($\psi$, system) while another subset is ignored ($\xi$, environment). An individual coarse-grained history is described by the path $\psi^\alpha(t)$ along with all possible paths $\xi^\alpha(t)$.

When the probability of each history can be assigned individually, the system behaves like a classical one and we say it has decohered. This means that the quantum interference between any two elements of this set of histories is negligible and the probability of reaching the same final state through two different stories is the sum of the probabilities of each history. The interest in finding histories that have undergone decoherence lies in the fact that these histories will be the ones that describe the classical domains.

One way to measure the decoherence suffered by two histories is through the
decoherence functional ($D$), which is \cite{Gell-Har,Har}

\begin{eqnarray}
\fl &D(\alpha,\alpha')=\int_\alpha D\psi^1 \int_{\alpha'}D\psi^2
\;\delta(\psi_f-\psi'_f)
e^{iS_0(\psi^1)}\rho_s(\psi_i,\psi'_i)e^{-iS_0(\psi^2)} \nonumber\\
 \fl \int d\xi_i \; d\xi'_i\int^{\xi^1}_{\xi_i}D\xi^1
&\int^{\xi^2}_{\xi'_i}D\xi^2\;\delta(\xi^1-\xi^2)\;
e^{i[S_E(\xi^1)+S_I(\psi^1,\xi^1)]}\rho_E(\xi_i,\xi'_i)e^{-i[
S_E(\xi^2)+S_I(\psi^2,\xi^2)]}
\end{eqnarray}

where $S_0$ is the free action of the system, $S_E$ is the action of the environment, $S_I$ gives the interaction between the system and the environment and $\rho_0$ and $\rho_E$ are the system and environment density matrices respectively. It is assumed that the system and environment are initially uncorrelated and therefore the density matrix can be factorized.

The influence functional ($F$) is obtained through the integration of two final states of the environment that are the same, ie $\xi^1=\xi^2=\xi$ \cite{CalHu,Fey-Hib}

\begin{eqnarray}
\fl F(\psi^1,\psi^2)&=e^{iS_{IF}} \nonumber\\
\fl &=\int d\xi \int d\xi_i \; d\xi'_i
\int^{\xi}_{\xi_i}D\xi^1\int^{\xi}_{\xi'_i}D\xi^2
 e^{i[S_E(\xi^1)+S_I(\psi^1,\xi^1)]}\rho_E(\xi_i,\xi'_i)e^{-i[S_E(\xi^2)+S_I(\psi^2,\xi^2)]}
\end{eqnarray}  
Therefore, the decoherence functional is

\begin{equation}
\fl D(\alpha,\alpha')=\int_\alpha D\psi^1\int_{\alpha'}D\psi^2\;\delta(\psi_f-\psi'_f)
 e^{iS_0(\psi^1)}\rho_s(\psi_i,\psi'_i)e^{-iS_0(\psi^2)}\;e^{iS_{IF}(\psi^1,\psi^2)}
\end{equation} 
The weak decoherence condition is recovered when\cite{Gell-Har,Har}

\begin{equation}
 e^{-Im[S_{IF}(\psi^1,\psi^2)]}<<1\;\;\;\Rightarrow\;\;\;Im[S_{IF}(\psi^1,
\psi^2)]>>1
\end{equation} 
If the interaction between system and environment can be written by a current coupling as

\begin{equation}
 S_I(\psi,\xi)=\int d^4x \;J(\psi(x))\xi(x)
\end{equation}
and the environment corresponds to free fields, then the influence functional can be written in terms of Jordan and Hadamard propagators as \cite{CalHu}

\begin{eqnarray}
\fl S_{IF}(\psi^1,\psi^2)&=\frac{i}{4}\int d^4xd^4x' \left[J(\psi^1)-J(\psi^2)
\right](x) \left[J(\psi^1)+J(\psi^2) \right](x') G(x,x')+ \nonumber\\
&+\frac{i}{4}\int d^4xd^4x'\left[J(\psi^1)-J(\psi^2)\right](x) \left[J(\psi^1)-J(\psi^2)\right](x')G_1(x,x')
\end{eqnarray}
Since the currents $J(\psi)$ are real, all we need to consider to find the real part of the decoherence functional are propagators: the Jordan propagator ($G$) is imaginary while the Hadamard propagator ($G_1$) is real. Considering the factor $i$ before the influence functional, the imaginary part can be written as,

\begin{equation}
\fl Im(S_{IF})= \frac{1}{4}\int d^4x\;d^4x' \left[ J(\psi^1(x))-J(\psi^2(x))\right] 
\left[ J(\psi^1(x'))-J(\psi^2(x')) \right]  G_1(x,x') \label{fd}
\end{equation}

This is the expression to be computed to determine the decoherence of the scalar perturbations during inflation. The coupling current between the graviton and scalar fluctuation is given by (\ref{cc}). In the next subsection  we calculate this expression. But before that we rewrite the decoherence functional using the results of Section \ref{spCBR} (to relate the scalar perturbation whit the CBR temperature) in order to put the decoherence functional in terms of an observable.

\subsection{Decoherence functional computation} \label{DecFunComp}

The Hartle Gell-Mann formalism lets us choose the histories involved in the decoherence functional. In this work we wish to choose histories representing different CBR temperature outcomes. Since nonlinear effects are small, the CBR temperature is determined by the scalar perturbations, and these evolve as a nearly free field. Therefore we assume histories where the single gauge invariant scalar perturbation $\phi(\eta)$  (defined in the \textit{Newtonian gauge}) evolves as a free perturbation (as described in Section \ref{spCBR}) while tensor perturbations are totally unspecified.

We start the decoherence functional computation by writing equation (\ref{fd}) in terms of the coupling current given by (\ref{cc}), the Hadamard propagator given by equation (\ref{hp}) and the polarization tensors of the gravitons given by equation (\ref{tp}). The decoherence functional results in

\bea
\mathrm{Im}\left(S_{IF}\right)&=&\frac{81}{16}\epsilon^4\frac{m_{pl}^2}{H^2}\int\frac{d^3p}{\left(2\pi\right)^3}\frac{d^3q}{\left(2\pi\right)^3}
\frac{d^3p'}{\left(2\pi\right)^3}\frac{d^3q'}{\left(2\pi\right)^3}\left(2\pi\right)^3\delta\left(\mathbf{p+q+p'+q'}\right)\nn
&&\Delta_{\left(\mathbf{p+q}\right)ijlm}p_iq_jp'_lq'_m\nn
&&\int d\eta d\eta'\:\frac1{\left|\mathbf{p+q}\right|}G_{1\left|\mathbf{p+q}\right|}\left(\eta,\eta'\right)\eta^{-1}
\eta'^{-1}F_p\left(\eta\right)F_q\left(\eta\right)F_{p'}\left(\eta'\right)F_{q'}\left(\eta'\right)\nn
&&\frac1{T_0^4}\left[\delta T_{\mathbf{p}}^1\delta T_{\mathbf{q}}^1-\delta T_{\mathbf{p}}^2\delta T_{\mathbf{q}}^2\right]\left[\delta T_{\mathbf{p'}}^1\delta T_{\mathbf{q'}}^1-\delta T_{\mathbf{p'}}^2\delta T_{\mathbf{q'}}^2\right]
\label{aleph}
\tea
Let us assume

\be
\frac1{T_0^2}\left\langle \delta T_{\mathbf{p}}^a\delta T_{\mathbf{q}}^b\right\rangle=\frac N{p^3}\delta^{ab}\delta\left(\mathbf{p+q}\right)
\label{a1}
\te
where $N\approx 10^{-10}$ is the square of the fractional temperature fluctuation of the Cosmic Background Radiation given by the current observations \cite{Wei} and the regularization $\left.\delta\left(\mathbf{k}\right)\right|_{\mathbf{k}=0}=L_0^3$.

Then

\bea
\left\langle \mathrm{Im}\left(S_{IF}\right)\right\rangle&=&\frac{81}{4\left(2\pi\right)^3}\epsilon^4\frac{m_{pl}^2N^2}{H^2}L_0^3\int\frac{d^3p}{\left(2\pi\right)^3}\frac{d^3q}{\left(2\pi\right)^3}
\frac{\Delta_{\left(\mathbf{p+q}\right)ijlm}p_iq_jp_lq_m}{\left|\mathbf{p+q}\right|p^3q^3}\nn
&&\int \frac{d\eta}{\eta} \frac{d\eta'}{\eta'}\:G_{1\left|\mathbf{p+q}\right|}\left(\eta,\eta'\right)
F_p\left(\eta\right)F_q\left(\eta\right)F_{p}\left(\eta'\right)F_{q}\left(\eta'\right)
\label{aleph2}
\tea
where $\Delta_{\left(\mathbf{p+q}\right)ijlm}p_iq_jp_lq_m=4\left(\mathbf{p}\times\mathbf{q}\right)^2$.

The conformal time integrals may be written as a sum of two squares, $I_1^2+I_2^2$, where
\numparts
\be
I_1=\int \frac{d\eta}{\eta} \left[\cos \left(k\eta\right)+\frac{\sin\left(k\eta\right)}{k\eta}\right]F_p\left(\eta\right)F_q\left(\eta\right)
\label{aleph5}
\te

\be
I_2=\int \frac{d\eta}{\eta} \left[\frac{\cos\left(k\eta\right)}{k\eta}-\sin \left(k\eta\right)\right]F_p\left(\eta\right)F_q\left(\eta\right)
\label{aleph6}
\te
\endnumparts
where $k=\left|\mathbf{p+q}\right|$. 
Now

\bea
\frac1{\eta}F_p\left(\eta\right)F_q\left(\eta\right)&=&\frac{pq\eta}2\left[\cos\left(p-q\right)\eta-\cos\left(p+q\right)\right]\nn
&+&\frac{p}{2}\left[\sin\left(p+q\right)\eta+\sin\left(p-q\right)\eta\right]\nn
&+&\frac{q}{2}\left[\sin\left(p+q\right)\eta-\sin\left(p-q\right)\eta\right]\nn
&+&\frac1{2\eta}\left[\cos\left(p+q\right)\eta+\cos\left(p-q\right)\eta\right]
\label{aleph7}
\tea

Keeping only the large $\left|\eta\right|$ terms is consistent with assuming that most decoherence happens when modes are within the horizon. Keeping only the highest powers in conformal time,  we transform this integral into

\begin{equation}
\fl \left\langle \mathrm{Im}\left(S_{IF}\right)\right\rangle=\frac{81}{\left(2\pi\right)^3}\epsilon^4\frac{m_{pl}^2N^2}{H^2}L_0^3\int\frac{d^3p}{\left(2\pi\right)^3}\frac{d^3q}{\left(2\pi\right)^3}
\frac{\left(\mathbf{p}\times\mathbf{q}\right)^2}{\left|\mathbf{p+q}\right|pq}\int {d\eta} {d\eta'}\eta\eta'\cos\left\{\Omega_{\mathbf{pq}}\left(\eta-\eta'\right)\right\}
\label{aleph3}
\end{equation}
where $\Omega_{\mathbf{pq}}=\left|\mathbf{p+q}\right|-p-q$.

The time integrals may be performed to yield

\begin{equation}
 \left\langle \mathrm{Im}\left(S_{IF}\right)\right\rangle=\frac{4\: 81}{\left(2\pi\right)^3}\epsilon^4\frac{m_{pl}^2N^2}{H^2}L_0^3\int\frac{d^3p}{\left(2\pi\right)^3}\frac{d^3q}{\left(2\pi\right)^3}
\frac{\left(\mathbf{p}\times\mathbf{q}\right)^2}{pq\left|\mathbf{p+q}\right|}
\frac{\vert \eta_i \vert^2} {\Omega_{\mathbf{pq}}^2} \sin^2\left(\Omega_{\mathbf{pq}}\eta_i\right)
\label{aleph4}
\end{equation}

Observe that the integrand is well behaved as $p,q\to 0$, so we can extend the momentum integrals all the way to the origin. Also because of the large frequency involved, we may approximate the $\sin^2$ by $1/2$. The only dimensioned quantity which remains is the upper integration limit $\left|\eta_r\right|^{-1}$, and we get by dimensional analysis

\be
\left\langle \mathrm{Im}\left(S_{IF}\right)\right\rangle\approx\epsilon^4\frac{m_{pl}^2N^2}{H^2}L_0^3\frac{\left|\eta_i\right|^2}{\left|\eta_r\right|^5}
\label{aleph8}
\te
which is about $10^{120}\epsilon^4$.

\subsection{Quadratic deviation of the influence functional}
If we regard $\mathrm{Im}S_{IF}$ as a stochastic variable, we may devise a Feynman graph representation for its moments. These graphs are composed of graviton lines and CBR lines joining at cubic vertices, according to the Feynman rules

a) a graviton line carries a momentum label $\mathbf{k}$ and coordinate labels $ij$ and $lm$ and time labels $\eta$ and $\eta'$ at each end. It corresponds to the element

\be
\frac1{a(\eta)a(\eta')m_{pl}^2}\frac{\Delta_{\left(\mathbf{k}\right)ijlm}}{\left|\mathbf{k}\right|}G_{1\left|\mathbf{k}\right|}\left(\eta,\eta'\right)
\label{aleph9}
\te
b) a CBR line carries a momentum $\mathbf{p}$ label and also labels  time $\eta,\eta'$ and ''history'' $a,b$ at each end. It corresponds to the element

\be
\frac {N\epsilon^2}{p^3}\delta^{ab}F_p\left(\eta\right)F_{p}\left(\eta'\right)
\label{aleph10}
\te
c) a vertex joins a graviton line (labels $\mathbf{k},ij,\eta$), a CBR line (labels $\mathbf{p},i,\eta,a$) and a second CBR line (labels $\mathbf{q},j,\eta,b$). It corresponds to the element

\be
\frac{m_{pl}^2}{2H^2\eta^2}p^iq^j\sigma^{ab}_3\delta\left(\mathbf{p+q+k}\right)
\label{aleph11}
\te
(for an outgoing line the sign of momentum is reversed). $\sigma_3$ is the third Pauli matrix $diag\left(1,-1\right)$. Observe that tadpoles vanish identically because of the sum over the history label.

In this language, $\left\langle\mathrm{Im}S_{IF} \right\rangle$ is the setting sun graph \cite{CalHu}. The second moment $\left\langle \left(\mathrm{Im}S_{IF}\right)^2\right\rangle$ corresponds to graphs containing two graviton lines and four vertices. We discard graphs containing tadpoles and also the disconnected graph, which equals $\left\langle\mathrm{Im}S_{IF} \right\rangle^2$. The remaining graphs contain three loops and therefore four CBR lines. Since they are connected, there is a single overall delta function from momentum conservation which contributes a factor of $L_0^3$ to the final amplitude. From simple power counting, we get

\be
\left\langle \left(\mathrm{Im}S_{IF} \right)^2\right\rangle-\left\langle\mathrm{Im}S_{IF} \right\rangle^2\approx\left(\frac{m_{pl}^2}{H^2}\right)^4\left(\frac1{m_{pl}^2}\right)^2\left(N\epsilon^2\right)^4L_0^3H^4J
\label{aleph12}
\te
where $J$ is the remaining momentum and time integration. We analyze this in the same terms as in the previous section to conclude that $J\propto \left|\eta_i\right|^{4}\left|\eta_r\right|^{-7}$. We therefore find

\be
\frac{\left\langle \left(\mathrm{Im}S_{IF} \right)^2\right\rangle-\left\langle\mathrm{Im}S_{IF} \right\rangle^2}{\left\langle\mathrm{Im}S_{IF} \right\rangle^2}\propto\left(\frac{\left|\eta_r\right|}{\left|\eta_i\right|}\right)^3\approx e^{-180}
\label{aleph14}
\te
This result, together with the result for $\langle Im(S_{IF}) \rangle$, shows that the decoherence functional behaves as a gaussian variable strongly centered around its mean value. Furthemore, this mean value is large enough to produce an effective decoherence process on the scalar perturbation, making it impossible to detect quantum effects on the CBR spectrum.

\section{Conclusions} \label{conc}

In this paper we have computed the decoherence functional between two histories of the Universe where scalar primordial fluctuations evolve in a prescribed way while tensor fluctuations are regarded as an environment. This decoherence functional is relevant to the question of whether it is possible to detect nontrivial quantum behavior in observations of the CBR temperature alone (that is, blind to CBR polarization). Our result implies that such detection is unrealistic by today's standards. Because of the well known triangle inequality \cite{AraLie70,Lub78,LubLub93,LloPag88,Pag93}, we expect the same would be obtained if the scalar fluctuations were regarded as an environment for the tensor ones.

This finding is consistent with earlier analysis of decoherence of cosmological fluctuations \cite{GP,CalHu95,CalGon97,KPS,KLPS,KP,Lom-Lop,ProRig07,Bur,Mar}. We hold this paper as an advance with respect to those earlier analysis because our system-environment split is related to the features of a realistic observational scheme, rather than just being assumed. Moreover, we make no ad-hoc assumptions regarding the model, since the only coupling we are considering is demanded by general relativity. The present work is probably closest to \cite{CalHu95}, but goes beyond it in that the proper gauge invariant degree of freedom is identified, rather than just the inflaton field.
 
Finding tangible proof of the quantum nature of our Universe is one of the most fascinating challenges faced by Cosmology today. We believe our result should not be read in a negative way but rather in a positive one, as pointing to the direction in which a successful scheme could be found. We are continuing our research with this goal in mind.

\section*{Acknowledgments}
This work is supported by the University of Buenos Aires, CONICET and ANPCyT. We acknowledge valuable discussions with D. Lopez Nacir, F. Lombardo and F. D. Mazzitelli.

\end{document}